\pdfoutput=1
\documentclass[10pt,a4paper]{article}
\RequirePackage{latexsym,amsmath,amssymb}
\RequirePackage[dvipsnames,usenames]{color}

\usepackage{fullpage}
\usepackage{cancel}
\usepackage[british]{babel}
\usepackage[latin1]{inputenc}
\usepackage[T1]{fontenc}
\usepackage[final]{showkeys} 
\usepackage{hyperref}
\hypersetup{pdfauthor={M. Cadoni, A. Sanna...},
            pdftitle={}
            }
\usepackage[]{cleveref}
\Crefname{equation}{Eq.~}{Eqs.}
\Crefname{figure}{Fig.}{Figs.}
\crefformat{plural}{#2eqs.~(#1)#3}
\crefname{section}{Sect.}{Sects.}

\usepackage{amsfonts}
\usepackage{graphicx}
\def\lb{\label}
\def\be#1\ee{\begin{align}#1\end{align}}

\def\lb{\label}

\title{\bf  Anisotropic Fluid  Cosmology: an Alternative to Dark Matter? }
\author{M.~Cadoni${}^{ab}$\thanks{E-mail: mariano.cadoni@ca.infn.it},
A. P. ~Sanna${}$\thanks{E-mail: asanna9564@yahoo.it} \ 
and
M. ~Tuveri${}^{b}$\thanks{E-mail: matteo.tuveri@ca.infn.it}
\\
\\
${}^a$\emph{Dipartimento di Fisica, Universit\`a di Cagliari}
\\
{\em Cittadella Universitaria, 09042 Monserrato, Italy}
\\
\\
${}^b$\emph{I.N.F.N, Sezione di Cagliari, Cittadella Universitaria, 09042 Monserrato, Italy}
\\
\\}
\begin{document}
\maketitle
\begin{abstract}
We use anisotropic fluid cosmology to describe the present, dark energy-dominated, universe.  Similarly to what has been proposed for galactic dynamics \cite{Cadoni:2017evg}, the anisotropic  fluid gives an effective description of baryonic matter, dark energy and their possible interaction, without  assuming the presence of dark matter. The resulting anisotropic fluid spacetime naturally generates inhomogeneities at small scales, triggered by an anisotropic stress, and could therefore be responsible for structure formation at these scales. Solving the cosmological equations, we show that the dynamics of the scale factor $a$ is described by usual FLRW cosmology  and decouples completely  from that describing  inhomogeneities. We assume that the cosmological anisotropic fluid  inherits   the  equation of state from that used in Ref. \cite{Cadoni:2017evg}  to describe galaxy rotation curves. We show that, in the large scale regime, the fluid can be described as a generalized Chaplygin gas and fits  well the distance modulus experimental data of type Ia supernovae, thus correctly modelling the observed accelerated expansion of the universe.  Conversely, in the small scale regime, we use cosmological perturbation theory to derive the  power spectrum  $P(k)$ for mass density distribution. At short wavelengths, we find a $1/k^4$  behaviour, in good accordance with  the observed correlation function for matter distribution at small scales.

\end{abstract}

\section{Introduction}
\lb{sec1}

Our  present understanding of cosmology, large scale structure of our universe and galactic dynamics is based on the $\Lambda$CDM model \cite{Aghanim:2018eyx}. This model  explains, in good agreement with observational data,  the present accelerated expansion of the universe  \cite{Riess:1998cb,Perlmutter:1998np}, cosmic microwave background observations,  structure formation, galaxy rotation curves  and gravitational lensing effects \cite{Rubin1980,Persic:1995ru, Massey:2010hh}. However,  the $\Lambda$CDM model is not completely satisfactory from a conceptual point of view. It postulates that about 95 percent of the matter contained in our universe is of exotic nature. At galactic level, it fails to explain the Tully-Fisher (TF) relation $v^2\sim \sqrt{a_0 G m_B}$ , which  relates the asymptotic velocity $v$ of stars in galaxies  to the galactic baryonic mass content $m_B$  and  to  $a_0$, an acceleration parameter of the same order of magnitude of the current value of the Hubble constant $H_0$ \footnote {Throughout this paper, we will mainly use natural units $c=\hbar=1$}. Moreover, there is also some tension between the $\Lambda$CDM model and observations at the level of galaxies,  galaxy clusters \cite{DelPopolo:2016emo,Kroupa:2012qj} and the determination of the Hubble parameter \cite{Aghanim:2018eyx, Riess:2019cxk, Aghanim:2019ame, VerdeHubble}.

Motivated mainly by the conceptual difficulties at the level of galactic dynamics, recently several alternative proposals have been put forward  to explain the galactic phenomenology  commonly attributed to dark matter \cite{Cadoni:2017evg, Verlinde:2016toy,Cadoni:2018dnd,Tuveri:2019zor,Smolin:2017kkb}. Typically,  these alternative approaches use infrared modifications of general relativity (GR). Efforts along this direction have been undertaken in the emergent gravity scenario \cite{Verlinde:2016toy}, where the  additional force at galactic level is  generated by the interaction between baryonic matter and dark energy (DE), in the corpuscular gravity scenario \cite{Cadoni:2017evg,Cadoni:2018dnd, Tuveri:2019zor} and  in  approaches which assume  an environmental modification of the inertial/gravitational mass ratio \cite{Smolin:2017kkb,Alexander:2018lno}.
A common feature of  these attempts is the fact that, in the weak field regime, they all reproduce Milgrom's MOdified Newtonian Dynamics (MOND) \cite{Milgrom:1983ca, Milgrom:2014usa}, which gives a simple explanation of the Tully-Fisher relation and promotes $a_0$ to a fundamental constant \cite{Milgrom1983}.

So far, the previously mentioned attempts have been mainly confined to galactic dynamics. However, there are several reasons that strongly motivate  their extension to cosmology. Firstly, dark matter plays a crucial role not only in galactic dynamics, but also  in structure formation \cite{DelPopolo:2008mr}. Any alternative to dark matter should therefore not only explain anomalous galactic rotation curves, but also structure formation.  Secondly, the threshold acceleration parameter $a_0$, appearing in the TF relation, has the same order of magnitude of the Hubble constant $H_0$, indicating the existence of a deep connection between galactic dynamics and cosmology.
Last but not least, in the emergent gravity scenario, the additional force beyond the Newtonian one is a "dark force" originated from  the  response  of dark energy to the presence of  baryonic matter, linking again the physics at  galactic scales  to cosmology. 

It is therefore tempting  to look for a unified description  encompassing different regimes of gravity: Newtonian,   galactic,   cosmological. 
The origin of the dark force as baryonic matter-DE interaction clearly indicates that the first step along this direction should be the extension of the dark force idea to the cosmology of a dark energy-dominated universe.  This paper is devoted to the attempt of building  such a cosmological model, motivated by the emergent gravity description of galactic dynamics, without  assuming the presence of dark matter. It is known that, at galactic level, dark force effects allow for an effective description in terms of GR sourced by an anisotropic fluid \cite{Cadoni:2017evg}. We will therefore use anisotropic fluid cosmology as an effective description of the cosmological effects of DE-baryonic matter interaction in a  dark energy-dominated universe. The use of an anisotropic fluid as model for baryonic matter and DE is quite natural in this context because it is known that  an anisotropic fluid is equivalent to a two-fluid system \cite{Bayin1986, Zimdahl:1996fj}. 

The structure of this paper is as follows. In Sect. \ref{sec1a} we argue that an anisotropic fluid spacetime naturally encodes the presence of inhomogeneities at  small scales and can therefore be used to describe  cosmic structures at these scales.  In Sect.  \ref{sec2} we set up our cosmological model sourced by an anisotropic fluid  and solve the cosmological equations.  We show that the dynamics of the scale factor $a$ is described by usual  Friedmann-Lemaitre-Robertson-Walker (FLRW) cosmology and decouples completely  from that describing  inhomogeneities. We also  discuss observational constraints on the presence of inhomogeneities.  The equation of state (EoS) for our anisotropic fluid is described in Sect. \ref{seceos}. 
We assume that this EoS is  inherited from that used  in Ref. \cite{Cadoni:2017evg} to describe the effects of the dark force in galactic dynamics.

In Sect. \ref{seccg} we investigate  the large scale regime  of our cosmological model. We show that the predictions of our model fits very well the distance modulus data of type Ia supernovae, thus correctly describing the observed accelerated expansion of the universe. Moreover, we show that, in this regime, the fluid allows for a description in terms of a generalized Chapligyn gas. 
In Sect. \ref{secset} we rewrite the stress-energy tensor for the anisotropic fluid in a form suitable for treating anisotropic stress as a perturbation.
Cosmological perturbations in our model are described in Sect. \ref{sect4} as perturbations of the de Sitter (dS) background. We first consider isotropic perturbations, which  describe the behaviour of the power spectrum  $P(k)$ for mass density distribution at large wavelengths. Thereafter, we consider perturbations due to the anisotropic fluid, which  are described by an anisotropic  stress tensor. This allows us to derive the form of  $P(k)$ at small wavelengths. We find $P(k)\sim 1/k^4$, with  good accordance with the observed correlation function for matter distribution at small scales.
Finally, in Sect. \ref{concl} we present our conclusions.

\section{Inhomogeneous cosmology in the dark energy-dominated era}
\lb{sec1a}
At sufficiently large scales our universe appears to be homogeneous and isotropic. This is an observational fact, which has been codified in the cosmological principle,  the basis  of modern  cosmology.  On the other hand,  the existence of   structures implies that at smaller scales the universe is inhomogeneous and anisotropic. For what concerns inhomogeneity, on which this paper is mainly focused, the transition scale is about  $R=100-300 \ \text{Mpc}/h$\, \cite{Hogg:2004vw,Yadav:2005vv,Yadav2010,Marinoni2012,Clowers2013,Ntelis:2016suu},  where $h$ parametrizes the Hubble constant, i.e. $H_0 = 100 h \ \text{km} \ \text{s}^{-1} \ \text{Mpc}$.  
Thus, if one focuses only on the  present, dark-energy dominated  era, the simplest  description of our universe  should be that of an inhomogeneous cosmological model, in which  inhomogeneities disappear when averaged  at scales larger than $R$. 

In order to make things as simple as possible, we assume that isotropy is preserved at the level of the metric and matter density. We will only allow for  anisotropies  in the fluid  pressure, in  the form of different values for its radial and perpendicular components. As we will see  later,  in the case of  an anisotropic fluid, an anisotropic stress can be used to generate  inhomogeneities. We can parametrize inhomogeneities by means of a function  of the radial coordinate $r$,    $\mathcal{E}(r)$, which goes to zero fast enough for $r>R$. In the limit $r\to \infty$, usual FLRW cosmology is recovered.

Although it is natural and simple,  this pattern  is not the way the standard model of cosmology - the $\Lambda$CDM model - uses to describe inhomogeneities  and structure formation. In the $\Lambda$CDM framework, they are explained in terms of the gravitational growth due to dark matter   of primeval scale-invariant and gaussian  perturbations generated during inflation \cite{DelPopolo:2008mr, Blumenthal1984}.  

The early universe is homogeneous and isotropic, the departures from homogeneity at high redshift  are well described by perturbation theory, which results in a scale-invariant power spectrum for the mass distribution at long wavelengths 
\be\lb{ps}
P(k)=\langle |\delta(k,t)|^2\rangle \sim k,
\ee
where $\delta(k,t)$ is the Fourier transform of mass density contrast at  wave-vector $k$.
At smaller redshifts, $z\lesssim10^4$, radiation pressure and the dynamics of non relativistic matter produce a bend in the power spectrum, which is described by a transfer  function $T(k)$ \cite{Peebles:2002gy, AmendolaDE}:%
\be\lb{ps1}
P(k)\propto k\, T^2(k).
\ee
$T(k)$ depends on  the details of cosmological model we are using and on damping and dissipation effects.  More precisely,  it is determined by the dark matter model, its interaction with the other cosmological fluids (pressureless matter, collisional photons, collisionless neutrinos, etc.) and their relative densities.  $T(k)$  must be therefore determined by numerically solving the Boltzmann equation in an expanding background \cite{Bardeen1986,Bond83, Hu97}.
At the epoch in which matter and radiation densities are equal,  assuming also a cold dark matter model, the transfer function appears to be well described by the fit \cite{AmendolaDE, Bardeen1986}
\be
T(k) = \frac{\ln(1+0.171x)}{0.171 x}\left[1+0.284x+(1.18x)^2+(0.399x)^3+(0.490x)^4 \right]^{-1/4},
\lb{fitform}
\ee
where $x\equiv k/k_{eq}$, with $k_{eq}$ characterizing the wavenumber of Fourier modes at the equivalence epoch. The fit \ref{fitform} is engineered to reproduce the inflationary scale-invariance power spectrum at large scales,  since, on super-horizon scales  matter perturbations are frozen and there is no damping process to alter the primordial power spectrum. In this regime, in fact, which corresponds to the limit $x \ll 1$, the transfer function behaves approximately as $T(k) \sim 1$, yielding the inflationary power spectrum (\ref{ps}). On small scales, i.e. $x \gg 1$, the transfer function has a $k$-dependence, $T(k) \sim \ln k/k^2$, and thus the power spectrum (\ref{ps1}) approximately goes as \cite{Peebles:2002gy}:
\be \lb{ps1.2}
P(k) \sim k^{-3}.
\ee
Another method to describe the statistical distribution of cosmic structures is given by the two-point correlation function, $\xi(r)$, which quantifies the probability of finding two structures, separated by a distance $r$, in excess with respect to a random background distribution. A well-known result is that the two-point correlation function is related to the power spectrum, being its Fourier transform: 
\be
P(k) = \frac{1}{k^3} \int_0^{\infty} \xi(r) \frac{\sin(kr)}{kr}4\pi r^2 \ dr.
\label{Ps2corr}
\ee
Observations show that, at physical scales ranging from $100 \ \text{kpc}/h$ to  $10  \ \text{Mpc}/h$, $\xi(r)$ is well-described by a simple power-law  \cite{LongairCosmology, Totsuj1969, Peebles1, Peebles2, Peebles:2001cy} 
\be
\xi(r) = \left(\frac{r}{r_0} \right)^{-\gamma},
\label{corrpowerlaw}
\ee 
where $r_0 \sim 5 \ \text{Mpc}/h$ is the so-called ``correlation length'', at which the probability of finding two galaxies, at a given distance $r$ from each other, is greater than the background one by a factor of 2, while $\gamma$ is experimentally determined. Observations suggest that $\gamma \in \left[1.8, 2\right]$ \cite{LongairCosmology, Totsuj1969, Peebles1, Peebles2, Peebles:2001cy}. Plugging Eq.~(\ref{corrpowerlaw}) into Eq.~(\ref{Ps2corr})  and using $\gamma = 2$ yields:
\be
P(k)  \sim k^{-4}.
\label{ps2}
\ee
At scales larger than  $10 \ \text{Mpc}/h$, the correlation function decreases more rapidly than the power-law (\ref{corrpowerlaw}) \cite{LongairCosmology,Peebles:2001cy, Zehavi2004}. This behaviour is thought to be due to the fact that, at larger separation distances (smaller $k$), the galaxy distribution becomes anticorrelated, i.e. $\xi < 0$, yielding a power spectrum which increases with increasing wave-number \cite{Peebles:2001cy}. In the context of the $\Lambda$CDM model, these departures from the power-law behaviour encode information about the relation between galaxies and dark matter halos \cite{Zehavi2004}.\\

It is quite evident that an inhomogeneous cosmological model, characterized by the  function $\mathcal{E}(r)$, does not represent a full alternative to the $\Lambda$CDM model. It cannot be proposed as a model describing the full history of our universe, in particular  the early universe. This is because mass distribution is intrinsically gravitationally unstable in FLRW cosmology. The only viable way to explain the scale-invariant large scale behaviour of the power spectrum  (\ref{ps}) is to assume that it has been generated by very small perturbations in the early universe described by inflationary cosmology. 

On the other hand, an inhomogeneous cosmological model, parametrized by the function $\mathcal{E}(r)$, can be used as an effective description of the present, dark energy-dominated era of our universe,  i.e.  for redshift  $z\lesssim 1$ \cite{Aghanim:2018eyx}. In particular, we expect this  inhomogeneous model to determine the short wavelength behaviour (\ref{ps2}), since the latter is valid on scales which are smaller than the homogeneity transition scale mentioned at the beginning of this section.  

At galactic scales, gravity sourced by an anisotropic fluid can give an  effective description of the additional force commonly attributed to dark matter \cite{Cadoni:2017evg}. Since, in the $\Lambda$CDM model, dark matter plays a crucial role for structure formation, the most natural candidate for the source in our inhomogeneous cosmology model  is that of an anisotropic fluid.

\section{Anisotropic fluid spacetime}
\lb{sec2}
As explained in the introduction, an anisotropic fluid can be used as a description of a two-fluid  model of baryonic matter, dark energy and their interaction. Moreover, as elucidated in the previous section, it is a promising candidate for describing the transition from an inhomogeneous universe at short scales to a homogeneous one at large scales, during the dark energy-dominated epoch.

Let us therefore set up  a cosmological model in which the various forms of matter, sourcing cosmological evolution and structure formation, are described  by an anisotropic fluid with energy-momentum tensor given by \cite{Cosenza1981, Herrera1997}

\begin{equation}
T_{\mu\nu} = \left(\rho + p_{\perp} \right)u_{\mu}u_{\nu} + p_{\perp} \ g_{\mu\nu} - \left(p_{\perp}-p_{\parallel} \right)w_{\mu}w_{\nu},
\label{TensoreEI}
\end{equation}
where the  fluid velocity  $u_{\mu}$ and the spacelike  vector $w_{\nu}$ satisfy  $ u^{\nu}u_{\nu} = -1$, $ w^{\nu}w_{\nu} = 1$ and  $u^{\mu}w_{\mu} = 0$.  The energy density is given by $\rho$ and $p_{\perp}, \ p_{\parallel}$ are, respectively, the pressures perpendicular and parallel to the spacelike   vector $w_{\nu}$. 

If we take  $p_{\perp}= p_{\parallel}$ and assume a spatially homogeneous and isotropic universe,  we get the usual FLRW cosmological model with $p, \ \rho$ and the scale factor of the metric depending on the cosmological time $T$ only. In this situation, cosmological evolution is sourced by  a perfect fluid with  equation of state $p=p(\rho)$, whereas the velocity field $u^{\nu}$ is free from rotation,  shear and acceleration.

The simplest way to achieve $p_{\perp}\neq  p_{\parallel}$, i.e. to  have a non trivial anisotropic fluid, is to allow for a dependence of  
$p_{\perp}, \ p_{\parallel}$ and $\rho$ from the radial coordinate $r$. The spacetime is not anymore homogenous, but remains  isotropic, the only manifestation of anisotropy being $p_{\perp}\neq  p_{\parallel}$, which therefore becomes the source of the inhomogeneities. This is  consistent with the  cosmological principle only if at large scales, i.e $r\to \infty$,  $p_{\perp}-  p_{\parallel}\to 0$, reinstating  homogeneity and isotropy of the solution. 

\subsection{Cosmological model}
\lb{seccosm}
A  convenient parametrization of the spacetime metric is   
\begin{equation}
ds^2 = a^2(t)\left[-f(r) e^{\gamma(r)}dt^2 + \frac{dr^2}{f(r)}+r^2 d\Omega^2\right]; \hspace{0.5 cm} d\Omega^2 = d\theta^2 + \sin^2 \theta \ d\phi^2,
\label{metrica}
\end{equation}
where $t$ is the conformal time, $a$ the scale factor and $f,\gamma$ are metric functions.
Choosing  an appropriate frame,   the fluid velocity vectors are given by $u^{\nu}= (a^{-1}f^{-1/2}e^{-\gamma/2},0,0,0),\,\,\,w^{\nu}=(0,a^{-1}f^{1/2},0,0,)$.  Einstein  equations  $R_{\mu \nu} -\frac{1}{2}g_{\mu\nu} R=G_{\mu\nu}= 8\pi G T_{\mu\nu}$ give three independent equations
\be
&3\left(\frac{\dot{a}}{a}\right)^2 -\frac{e^{\gamma}f}{r^2} \left(-1+f+rf' \right)= 8\pi G a^2 \ \rho f e^{\gamma};\label{E00}\\
&\frac{\dot{a}}{a f}\left(f'+f\gamma' \right)=0; \label{E0r}\\
&\frac{e^{-\gamma}}{r^2 a^2 f^2} \left[r^2 \dot{a}^2 + e^{\gamma} a^2 f \left(-1+f+rf'+rf\gamma' \right)-2r^2 a \ddot{a} \right]= 8\pi G \  p_{\parallel}  \frac{a^2}{f}, \label{Err}
\ee

where the dot and the prime  denote derivatives with respect to $t$ and $r$, respectively.

Covariant  conservation of the the stress-energy tensor gives two more equations:
\begin{equation}
\dot{\rho}+\frac{\dot{a}}{a}\left(3\rho+p_{\parallel} + 2p_{\perp} \right)=0,
\label{T0}
\end{equation}

\begin{equation}
p'_{\parallel} + \frac{2}{r}\left(p_{\parallel}-p_{\perp} \right) +\frac{1}{2}\left(\rho+p_{\parallel}\right)\left(\gamma'+ \frac{f'}{f}\right) = 0.
\label{Tr}
\end{equation}

The  form of the  spacetime metric (\ref{metrica}), together with Eq.~(\ref{TensoreEI}), describes, as particular cases, the various regimes of  gravity sourced by an (an)isotropic fluid: Newtonian, galactic, cosmological. 
When $\dot{a}=0$ (we  set $a=1$),  our model reproduces a  static, spherically symmetric, anisotropic fluid  space-time, which has been used for several application \cite{Cosenza1981, Herrera1997, Paul:2016nvi,Cho:2017nhx,Bharadwaj:2003iw,Faber:2005xc,Su2009}. In particular, it has been  used to explain galactic dynamics without assuming the presence of dark matter  \cite{Cadoni:2017evg}. In this latter case, the  radial pressure $p_{\parallel}$ gives an additional component to the acceleration (dark force) at galactic scales. This is what we call the MOND regime of gravity, because it reproduces the MOND theory in the weak-field approximation. If, in addition to $\dot{a}=0$,  we also impose $p_{\parallel}=p_{\perp}$, we obtain GR sourced by static, spherically symmetric perfect fluid (the Newtonian regime of gravity). \\
On the other hand, if $\dot{a}\neq 0$, our model describes a non-homogeneous cosmological model, which interpolates between the MOND regime at galactic scales and the usual  FLRW  cosmology at $r\to\infty$.

\subsection{Decoupling of cosmological degrees of freedom from inhomogeneities }
\lb{sec3}
When $\dot{a}\neq 0$,  Eq.~(\ref{E0r}) and  Eq.~(\ref{Tr}) can be solved, respectively, for $\gamma$ and $p_{\perp}$:
\be\lb{f1}
 e^{-\gamma} = f,\quad p_{\perp} = p_{\parallel}+\frac{r}{2}p'_{\parallel}. 
 \ee
 
 The remaining equations give then:

\begin{equation}
3 \left(\frac{\dot{a}}{a}\right)^2 + \frac{1-f-rf'}{r^2}= 8\pi G a^2 \rho;
\label{E003}
\end{equation}
\begin{equation}
\left(\frac{\dot{a}}{a}\right)^2-2\frac{\ddot{a}}{a} +\frac{f-1}{r^2} = 8\pi G a^2 p_{\parallel};
\label{Err3}
\end{equation}
\begin{equation}
\dot{\rho} + \frac{\dot{a}}{a} \left(3\rho + 3p_{\parallel} + rp'_{\parallel} \right) =0.
\label{T02}
\end{equation}

We have  three equations for the four variables $f(r), \ a(t), \ p_{\parallel}(r, t), \ \rho(r, t)$.  As usual in cosmology, the system has to be closed imposing an equation of state for the fluid, $p_{\parallel} = p_{\parallel}(\rho)$.

Using  Eqs.~(\ref{E003}), (\ref{Err3}), Eq.~(\ref{T02})  can be recast in the form 
\be
\partial_t \left(a^2 \rho \right) = \frac{3}{8\pi G}\frac{d}{dt} \left(\frac{\dot{a}^2}{a^2} \right),
\ee
which can  be easily integrated  to give
\be\lb{h1}
a^2 \rho(r,t) =  a^2(t) \hat \rho(t)\ + \mathcal{E}(r), \quad a^2\hat \rho(t)=\frac{3}{8\pi G} \mathcal{H}^2,
\ee
where $\mathcal{H} \equiv \dot{a}/a$ is  the  conformal Hubble parameter and  $\mathcal{E}(r)$ is an integration function, which depends on the radial coordinate $r$ only.   Physically, $\mathcal{E}(r)$ represents the inhomogeneities  in the baryonic matter density distribution. With this  result, we can now separate the $r$-dependent and the $t$-dependent parts in Eq.~(\ref{E003}).  The former determines the metric function  $f$
\be \lb{g1}
f= 1- \frac{2 G m_B(r)}{r}- \frac{2 G M}{r},
\ee
where $m_B(r)$ is the Misner-Sharp mass  associated with the inhomogeneities  in the baryonic matter
\be\lb{ms}
 m_B(r)=4\pi  \int dr \ r^2 \mathcal{E}(r),
 \ee
and $M$ is an integration constant with the dimensions of a mass. Using Eq.~(\ref{g1}) into (\ref{Err3}), we get:
\be\lb{f3}
a^2 p_{\parallel}(r,t)=  a^2 (t)\hat p(t)+ \mathcal{P}(r),\quad a^2\hat p(t)=\frac{1}{8\pi G }\left[\left(\frac{\dot{a}}{a} \right)^2 -2\frac{\ddot{a}}{a}\right],\quad \mathcal{P}(r)= -\frac{m_B(r)+M}{4\pi r^3}. 
\ee
Eqs.~(\ref{h1}) and (\ref{f3}) clearly show that the  contributions of inhomogeneities ($r$-dependent terms)  to matter density and pressure can be separated from the homogeneous ($t$-dependent) cosmological contributions. This, in turn,  allows us to separate the dynamics of cosmological evolution, which determines  $a,\hat \rho, \hat p$, from the effect of inhomogeneities. In fact Eqs.~(\ref{E003}), (\ref{Err3}) and Eq.~(\ref{T02})  are completely equivalent to  the FLRW equations for $a,\hat \rho, \hat p$,
\be\lb{j1}
a^2\hat \rho=\frac{3}{8\pi G} \left(\frac{\dot{a}}{a} \right)^2,\quad a^2\hat p=\frac{1}{8\pi G }\left[\left(\frac{\dot{a}}{a} \right)^2 -2\frac{\ddot{a}}{a}\right],\quad \dot{\hat \rho} + \frac{\dot{a}}{a} \left(3\hat \rho + 3\hat p \right) =0, 
\ee
together with
\be\lb{pre}
\mathcal{P}(r)= -\frac{m_B(r)+M}{4\pi r^3},
\ee
which  determines the pressure  $\mathcal{P}(r)$ from $m_B(r)$ given by Eq.~(\ref{ms}). The perpendicular component of the pressure $p_{\perp}$  is then  determined from $p_{\parallel}$
using Eq.~(\ref{f1}).

 This is a quite interesting result:  cosmological degrees of freedom decouple from inhomegenities. 
This implies that the scale factor $a$ is completely determined by the homogeneous and isotropic component of density and pressure $\hat \rho, \hat p$ through the usual FLRW equations (\ref{j1}), whereas the only effect of inhomogeneities is to produce  a non-vanishing, $r$-dependent,  pressure (\ref{pre}).

Taking into account also Eq.~(\ref{g1}), the physical interpretation of Eq.~(\ref{pre}) is quite simple. The term proportional to $M$ gives a Schwarzschild-like contribution, i.e. an inhomogeneity    generated by a point-like source located at $r=0$. Its presence is not compatible with  observations, we have therefore to set the integration constant $M=0$.

The term proportional to $m_B(r)$ gives instead the contribution of spherically symmetric inhomogeneities distributed with density $\mathcal{E}(r)$. Since we want to recover usual FLRW cosmology at large distance ($r\to \infty$) we  have to assume  $m_B(r)\sim  -\frac{1}{2G}\mathcal{K} r^3 + \frac{c_1}{r}$ with  $\mathcal{K}=0, \pm 1$.  As we shall see in detail in the next section, the first term gives the spatial curvature of the spacetime, whereas the second one gives a contribution to  $f$ and $p_{\parallel}(r,t)$ that vanishes in the $r\to \infty $ limit. The  physical effect   of the $\mathcal{P}(r)$ term in Eq.~(\ref{pre}) can be  explained as a Newtonian contribution to the  pressure, $P_N=  \frac{1}{4\pi}\frac{m_B}{r^3}$, which produces   the   radial acceleration $a^r= 4\pi Gr P_N$ \cite{Cadoni:2017evg}.

\subsection{FLRW cosmology}   
\lb{sec4}
Usual FLRW cosmology can be obtained as a limiting case of our anisotropic fluid cosmology in two different, albeit related, ways.
In the first way, standard cosmology is obtained in the large scale limit $r\to\infty$. In fact, in this limit, both $\mathcal{P}(r)$ and $\mathcal{E}(r)$ go to zero, $p_{\parallel}=p_{\perp}=\hat p$, $\rho=\hat\rho$ and  Eqs.~ (\ref{j1}) become the FLRW equations written in terms of $p_{\parallel}$ and $\rho$.
The  same equations can be obtained by setting the integration function $\mathcal{E}(r)$ identically to zero, so that we identically get $p_{\parallel}=p_{\perp}=\hat p$ and $\rho=\hat\rho$.

It is quite interesting to notice that the derivation of the FLRW equations as limiting case of anisotropic fluid cosmology allows us to generate the constant spatial curvature term in that equations from a constant contribution  to the  density function $\mathcal{E}(r)$.
In fact, setting  $\mathcal{E}(r)=-\frac{3}{8\pi G}\mathcal{K}$,  we get $m_B=-\frac{\mathcal{K}}{2} r^3$, $f=1+\mathcal{K}r^2$ and Eqs.~(\ref{j1}) become
\be \lb{z2}
3 \left(\frac{\dot{a}}{a}\right)^2 -3 \mathcal{K}= 8\pi G a^2  \rho\, \quad \left(\frac{\dot{a}}{a}\right)^2-2\frac{\ddot{a}}{a}+\mathcal{K} = 8\pi G a^2  p,\quad \dot{\rho}+\frac{\dot{a}}{a}\left(3 \rho+3 p\right)=0,
\ee
where, for notation simplicity, we  set $p_{\parallel}=p$. 

Although our model works also for a 3D space with constant positive, negative or zero curvature, in the following we will consider, consistently with observations, only FLRW cosmologies with   $\mathcal{K}=0$. 

In this paper, we use the simplest description of dark energy , i.e. that of a cosmological constant $\Lambda$, which corresponds to constant energy density $\rho = \Lambda/8\pi G$ and  equation of state $p=-\rho$. 
As it is well-known, in this case, Eqs.~(\ref{z2}) (with  $\mathcal{K}=0$) give as solution the dS spacetime:
\be\lb{ds}
a= \sqrt{\frac{\Lambda}{3}} \frac{1}{t}.
\ee
If we  write the solution using the  cosmological time $T$ instead of the conformal time $t$, we get $a= e^{\sqrt{\frac{\Lambda}{3}}T}$.

\subsection{Observational constraints}
\label{obsconstr}
The isotropy and homogeneity of the universe at large scales, encoded in the cosmological principle, are to be considered valid from a statistical point of view, since the universe appears inhomogeneous on small scales, due to the presence of cosmic structures. This necessarily introduces a smoothing scale above which the universe appears statistically homogenoeus and isotropic. This transition scale represents essentially the threshold, below which an inhomogeneous and/or anisotropic description of the universe can be considered observationally viable, and above which the cosmological principle must be restored.
In the following, we briefly discuss some recent results concerning both the transition scale for isotropy and homogeneity.\\
\textbf{Scale of Isotropy}\\
The study reported in \cite{Marinoni2012} uses a sample of $930000$ luminous red galaxies from the Sloan Digital Sky Survey (SDSS), over a field of view of $9380\ deg^2$ and redshift $0.22 < z< 5$.  Focusing on  angular distribution of the galaxies, it is shown that the scale $R_{iso}$,  above which the angular distribution of galaxies appears to become statistically isotropic \textit{for all observers} (automatically implying homogeneity) is approximately $R_{iso} \sim 150 \ \text{Mpc}/h$. This result is found to be consistent with the N-body simulations of large scale structures formation.

An independent investigation \cite{Sarkar:2018smv}, based on an information entropy approach, shows that  $R_{iso}$ is about $200 \ \text{Mpc}/h$. The analysis is carried out on photometric data of $784329$ galaxies (up to redshift $z \sim 0.2143$) and spectroscopic data of $180181$ galaxies (up to redshift $z \sim 0.1341$), always from the SDSS. This study also shows that, at small scales, the galaxy distribution is highly anisotropic, in agreement with N-body simulations.\\
\textbf{Scale of Homogeneity}\\
The study reported in \cite{Ntelis:2016suu} uses a sample of galaxies, taken from the BOSS (Baryon Oscillation Spectroscopic Survey) CMASS survey, at redshift $0.43 < z < 0.7$. It shows that the scale above which the universe becomes homogeneous is about $R \sim 64.3 \ \text{Mpc}/h$. This is a result confirmed also by the analysis carried out in \cite{Hogg:2004vw}: using the Luminous Red Galaxy (LRG) spectroscopic sample of the SDSS, considering the redshift interval $0.2 < z < 0.35$, the homogeneity scale is found to be about $70 \ \text{Mpc}/h$. Similar conclusions are also found in \cite{Yadav:2005vv}.
Other investigations give a somehow larger transition scale $R$ of about $200-300 \ \text{Mpc}/h$ \cite{Yadav2010,Marinoni2012,Clowers2013}.

These results about the scale at which  inhomogeneities become relevant in cosmology, put a constraint on the form of  the function $\mathcal{E}(r)$, which in our model describes inhomogeneities. In particular, $\mathcal{E}(r)$ should be negligibly small at small redshift when the size of the universe is much larger than the transition scale R. This tells us that $\mathcal{E}(r)$ can be consistently put to zero when describing, for example, the Hubble diagram of high-redshift Type Ia Supernovae.

\section{The equation of state}
\lb{seceos}
A crucial issue  of our cosmological model is the determination of the EoS  that the anisotropic fluid must satisfy in the different regimes we are considering in this paper. The equation of state is not  only important to determine the background solution, but also to describe density perturbations  around the background solution. 
In our approach, the anisotropic fluid is meant to give an effective description of  dark energy, baryonic matter and their interaction. Although we know very well the EoS for the (perfect) fluids describing pure, non interacting,  DE (we are modelling it as a cosmological constant, therefore we have $p=-\rho$) or pure, non interacting, stiff baryonic matter, $p=0$, presently, we do not have a direct  way to derive the EoS for the anisotropic fluid describing  the interacting case.

The best thing we can do is to use what we know about the anisotropic fluid at galactic scales to infer information about the EoS of the fluid at cosmological level.
At galactic scales the interaction between  dark energy and baryonic matter is described by a dark force, which manifests itself through the radial component of the pressure of the anisotropic fluid \cite{Cadoni:2017evg}
\be
\lb{press}
p_{\parallel}=p= \frac{1}{4\pi r^2} \sqrt{\frac{m_B(r)}{G L}},
\ee
where $m_B(r)$ is the baryonic matter distribution in the galaxy and  $L \sim H_0^{-1}$ is the size of the cosmological horizon.

We will assume that the EoS for our anisotropic fluid  is inherited  from the  expression for the pressure in Eq.~(\ref{press}) that is responsible for  the dark force at galactic scales as explained in Eq.~Ref. \cite{Cadoni:2017evg}.
It would be nice to use exactly the same expression given by   Eq.~(\ref{press})  in the cosmological context of a  DE-dominated universe. However,  a simple argument shows that this is not possible and that Eq.~(\ref{press}) has to be  slightly  modified. In fact,  in this regime we expect the contribution of baryonic matter to be completely negligible. The mass appearing  under the square root has to be therefore considered as a total effective  mass $m_E(r)$, which is the sum of the baryonic, $m_B(r)$, the  DE contribution $m_{\Lambda}(r)$ and an interaction term $m_I(r)$:  $m_E(r)=m_B(r)+m_{\Lambda}(r)+m_I(r)$. At galactic scales,  $m_{\Lambda}$  and $m_I$  can be neglected, we have $m_E(r)\sim m_B(r)$ and we get Eq.~(\ref{press}).  
Conversely, in the limit $r\sim L$,   both baryonic matter  and  its interaction with DE can be neglected. From $\rho_\Lambda\sim \frac{1}{GL^2}$ we get    $m_{\Lambda}\sim \frac{L}{G}$, so that equation (\ref{press}) gives $p\sim -\rho_\Lambda$, where we have taken into account that the pressure in Eq.~(\ref{press}) is negative. We will therefore promote Eq.~(\ref{press}) to an {\sl effective equation of state} relating the radial pressure  of our anisotropic fluid with the effective matter density $\rho_E$ generating the effective mass $m_E$:
\be
\lb{eeos}
p_{\parallel}= \frac{1}{\sqrt{4\pi}\, r^2} \sqrt{\frac{\int d^3x\rho_E(r)}{G L}}.
\ee
Notice that in our description the effective matter density $\rho_E$ is expected to mimic the effects of both dark  and baryonic matter in the $\Lambda$CDM model.

Let us now consider the large scale  regime of our cosmological model, when the interaction between DE and baryonic matter  cannot be neglected, hence the EoS is expected to deviate from the simple form $p\sim -\rho$.
In the large scale limit $r\rightarrow \infty$, the contribution of inhomogeneities to the density $\rho$ and to pressure $p_{\parallel}$ dies out. Cosmological evolution is therefore described by the FLRW equations (\ref{z2}) with $\mathcal{K}=0$.  On the other hand,  although the universe  is  dominated by dark energy   the interaction of the latter  with baryonic matter cannot be  completely neglected.
 Being $m_E(r)=  \frac{4\pi a^3} {3} \rho_E (t) r^3$ and taking into account that, at large distances, we have $r\sim L$, we get from (\ref{press}) the equation of state:
\be
p = \frac{1}{\sqrt{12\pi G}L}\sqrt{\rho}\ a^{3/2}.
\label{DarkForceEOS}
\ee

In Sect. \ref{dsan}, we will consider inhomogeneities as  density perturbations of the dS background, in oder to describe the short wavelength behaviour of the   power spectrum (\ref{ps2}). In order to do this we will  consider in Eq.~(\ref{eeos}) both $p$ and $\rho_E$ as  small perturbations of the (constant) pressure and energy density sourcing the dS spacetime.

\section{Large scale cosmological regime and generalized  Chaplygin gas  model}
\lb{seccg}
We consider now the large scale  regime of our cosmological model.  We have seen in the previous  section that in the large scale limit, $r\rightarrow \infty$, cosmological evolution is described by the FLRW equations (\ref{j1}) with an  effective EoS for our fluid  given 
by (\ref{DarkForceEOS}).
 
Using the  EoS  (\ref{DarkForceEOS}), the cosmological equations (\ref{j1}) give
\begin{equation}\lb{t1}
\dot{a}^2-2\ddot{a}a = \frac{\sqrt{2}}{L}\ \dot{a}\  a^{7/2},
\end{equation}
which can be easily integrated by defining the new variable $K=\frac{\dot a}{\sqrt a}$ and the new time $\tau= \int a^{5/2} dt$.
The solution of Eq.~(\ref{t1}) is then given, in implicit  form, in terms of the conformal time $t$:
  
\be
t = -\frac{\sqrt 2\,c_1^5}{H_0} \biggr\{\ln \left(\frac{1- c_1 \sqrt{a}} {1+ c_1 \sqrt{a}}\right) + \frac{1}{2}\ln \left(\frac{1- c_1 \sqrt{a}+c_1^2 a} {1+ c_1 \sqrt{a}+c_1^2 a}\right)
+ \sqrt{3} \left[\text{arctg}\left(\frac{1-2 c_1\sqrt{a}}{\sqrt{3}}\right)-  \text{arctg}\left(\frac{1+2 c_1\sqrt{a}}{\sqrt{3}}\right)\right] \biggl\},
\label{fattorediscala}
\ee
where  $c_1$  is an integration constant.\\

From the form of the metric (\ref{metrica}) one can easily derive the conformal time $t$ in terms of the cosmological time  $T$ and the luminous distance
$D_L$  
\begin{equation}\lb{ff}
t = \int \frac{dT}{a}, \quad D_L = \frac{1}{a}\int_T^{T_0} \frac{dT}{a},
\end{equation}
where  $T_0$  is the present cosmological time. 
From this latter equation  it follows: 
\begin{equation}\lb{pp}
D_L = \frac{1}{a}\int^{a=1}_{a} \frac{dT}{a},
\end{equation}
where we have normalized to $1$ the scale factor at present cosmological time: $a(T_0)=1$.

$D_L$ can be calculated using Eqs.~ (\ref{fattorediscala}) and substituted into the distance modulus 
\begin{equation}
m-M = 25+5\log_{10} \left(\frac{D_L}{Mpc} \right),
\label{distancemodulus}
\end{equation}
where $m$ and $M$ are the apparent and absolute magnitude, respectively.

We can now compare  the theoretical prediction for the distance modulus of our cosmological model with EoS (\ref{DarkForceEOS}), as function of the redshift $z$, with the observational data for the Type Ia Supernovae, taken from the Supernova Cosmology Project (SCP) Union 2.1 Compilation  \cite{Amanullah2010}. 
 The scale factor (\ref{fattorediscala}) contains an  integration constant $c_1$, which  enters in the  relationship between conformal time $t$ and cosmological time $T$. It can be fixed by fitting the  prediction of our model with the observational  data. The fit gives the result $c_1= 0.769$. 

\begin{figure}
\centering
\includegraphics[width= 11 cm, height = 11 cm,keepaspectratio]{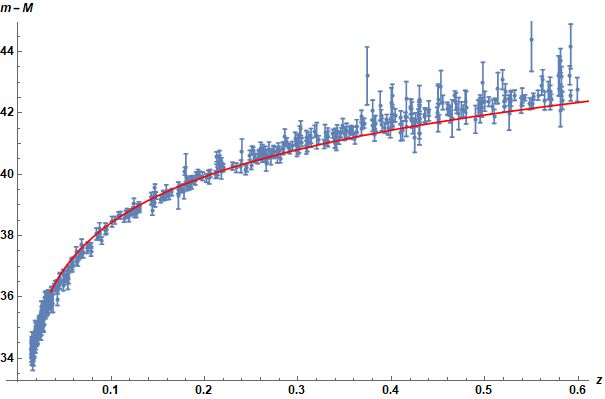}
\caption{Theoretical prediction of our cosmological model for the distance modulus as a function of the redshift (red line) vs  SNIa observational data, taken from Ref. \cite{Amanullah2010}. For the Hubble constant, we adopt the value $H_0 = 67.6 \ \text{km} \ \text{s}^{-1} \ \text{Mpc}^{-1}$, obtained combining Planck and BAO data \cite{Aghanim:2018eyx}. }
\label{Figure1}
\end{figure}

In Figure \ref{Figure1}, we show the comparison between the theoretical  prediction of our model for the distance modulus with the observations of \cite{Amanullah2010}, finding a good agreement. Notice that, in Fig. \ref{Figure1}, we have only considered observational data with $z \lesssim 0.6$. This corresponds to the range of validity of our cosmological model, which is meant to describe our late-time, dark energy-dominated universe.

\subsection{Connection with generalized Chaplygin gas model}
It is interesting to notice that our cosmological model  based on the EoS (\ref{DarkForceEOS}) belongs to the class of models termed  
new generalized  Chaplygin gas (NGCG) \cite{Zhang:2004gc,Wen:2017aaa}. This does not come as a surprise because these models are meant to give a unified description of dark energy and  dark matter.  This is  alike  to what we  have achieved by means of our anisotropic fluid cosmology,  but with an important difference. In our  description there is no dark matter, but only dark energy, baryonic matter and their interaction, whose effect at galactic scales and in cosmology should replace that of DM.
 
The EoS for NGCG has the general form $p_{NGCG} = \mathcal{A}(a)\rho_{NGCG}^{-\zeta}$.  More precisely,  it can be written as
\begin{equation}
p_{NGCG} = \frac{\zeta A \ a^{-3\left(1+\zeta\right)\left(1+\eta\right)}}{\rho_{NGCG}^{\zeta}},
\label{pressChap}
\end{equation}
where  $\zeta,\eta $  are some parameters and $A$ is a positive contant.

Comparing Eq.~(\ref{pressChap}) with our EoS  (\ref{DarkForceEOS}), we can determine the parameters $\zeta,\eta, A$:
\begin{equation}
\zeta = -1/2,\quad \eta=-2,\, \quad A=\frac{1}{3L}\sqrt{\frac{3}{\pi G}}= \frac{H_0}{3}\sqrt{\frac{3}{\pi G}}.
\end{equation}

\section{ The stress-energy tensor for anisotropic fluids}
\lb{secset}
In order to discuss cosmological perturbations, we need to rewrite the stress-energy tensor (\ref{TensoreEI}) for our anisotropic fluid 
as that pertaining to a  perfect fluid  plus a perturbation
\begin{equation}
T_{\mu\nu} = T^{(pf)}_{\mu\nu}+ \pi_{\mu\nu},\quad T^{(pf)}_{\mu\nu}=\left(\rho + P \right)u_{\mu} u_{\nu} + P g_{\mu\nu},
\label{TensEI}
\end{equation}
where $\pi_{\mu\nu}$ is the  anisotropic  stress tensor,  perturbing the stress-energy tensor for the perfect fluid $T^{(pf)}_{\mu\nu}$. \\
In general, the tensor $\pi_{\mu\nu}$ contains also dissipative contributions. We will neglect these contributions because heat flux vanishes  in the comoving   frame.
Comparing Eq.~(\ref{TensoreEI}) with Eq.~(\ref{TensEI}) one can easily read out the tensor $\pi_{\mu\nu}$
\begin{equation}
\pi_{\mu\nu} = \sqrt{3} \mathcal{S} \left[w_{\mu}w_{\nu} -\frac{1}{3}\left(u_{\mu}u_{\nu}+g_{\mu\nu} \right) \right],
\label{Anisotropicstress}
\end{equation}
where  $\mathcal{S}$ quantifies the degree of anisotropy:
\be\lb{fff}
\mathcal{S}= \frac{p_{\parallel}-p_{\perp}}{ \sqrt{3}}.
\ee
Using Eq.~(\ref{Anisotropicstress}) together with the equations $u^{\mu}u_{\mu} =-1,\, w^{\mu}w_{\mu} = 1,\,u^{\mu}w_{\mu} =0$ and going in the comoving frame,  one can easily check  that $\pi_{\mu\nu}$ satisfies the usual relations for a gauge-invariant anisotropic stress tensor:  $\pi^{\mu\nu}u_{\nu} = \pi^{\mu}_{\mu} =0$ and  $\pi_{00} = \pi_{0i}=0$.

In the comoving frame, the only non-vanishing components of $\pi^{\mu\nu}u_{\nu}$ are the spatial ones $\pi^{ij}${\footnote{ Spacetime indexes are denoted with greek letters,  space indexes  with latin letters: $\mu=(0,i)$.}. The spatial components of the stress-energy tensor are then
\be
T^1_1 =  P+\frac{2\mathcal{S}}{\sqrt{3}};\quad
T^2_2 = T^3_3 =  P-\frac{\mathcal{S}}{\sqrt{3}},
\label{comp2}
\ee
giving  for the radial and transverse component of the anisotropic fluid pressure:
\be
p_{\parallel} = P+\frac{2\mathcal{S}}{\sqrt{3}}; \label{pparallel},\quad
p_{\perp} =  P-\frac{\mathcal{S}}{\sqrt{3}}.
\ee
When $\mathcal{S}= 0$  we have  $p_{\parallel}=p_{\perp} =  P$, the fluid   is  perfect, homogeneous and isotropic. Conversely, $\mathcal{S}\neq 0$ implies $p_{\parallel}\neq p_{\perp}$,   signalizing anisotropic departure from a perfect fluid.

\section{Cosmological perturbations}
\lb{sect4}

In Sect. \ref{sec1a}, we have seen that one of the main goal of our anisotropic fluid cosmology is to describe structures at small  scales and in particular to derive the  matter power spectrum (\ref{ps2}). There are two different approaches  for doing  that.  
The first one is phenomenological: one just assumes the validity of our model and then finds the inhomogeneity function $\mathcal{E}(r)$ by fitting observational data about mass density distribution. Obviously, this approach has very low predictive power.

Alternatively, one can  consider $\mathcal{E}(r)$ as a small perturbation of a FLRW universe dominated by dark energy. Using  the simplest description for dark energy, that of a cosmological constant, we  need to consider perturbations near the dS cosmological solution generated by an anisotropic fluid. We will first  consider generic perturbations around a given cosmological background $g^{(0)}_{\mu\nu}$  given by Eq.~(\ref{metrica}) (we set $\gamma=0$ and $f=1$, i.e we consider a spatially-flat universe) in the linear regime. We will then specialize our calculations to the dS background.

We start from the usual form for the perturbed metric

\lb{sect5}
\begin{equation}
{g}_{\mu\nu}(t,x) =g^{(0)}_{\mu\nu}(t) + h_{\mu\nu}(t,x),
\label{metricaperturbata}
\end{equation}
where the background metric depends only on the conformal time $t$, whereas $h_{\mu\nu}(t,x)$ depends both on $t$ and on the spatial coordinates $x^i$. 

After some tedious  but straightforward calculation, one gets, for the perturbed components of the Einstein tensor $G_{\mu\nu}$ at the linear level in $h_{\mu\nu}$

\begin{equation}\begin{split}
\delta G_{00}  =&-3\ddot{a} \left(\frac{h_{00}}{a^3}+ah^{00} \right)+a\dot{a} \left(-2\partial_i h^{i0}-\frac{1}{2}\dot{h}^{ik}\delta_{ik}-6\mathcal{H}h^{00}-\frac{3}{2}\dot{h}^{00}\right)+\frac{1}{2a^2}\biggl(\mathcal{H} \delta^{ik}\dot{h}_{ik}-3\mathcal{H}\dot{h}_{00}+\\
&+\partial^i \partial^jh_{ij}-\delta^{ij}\partial^k \partial_k h_{ij} \biggr);
\label{dG00}
\end{split}\end{equation}

\begin{equation}\begin{split}
\delta G_{0i} & = \ddot{a}\left(a h^0_i -\frac{3h_{0i}}{a^3}\right)+ a\dot{a} \left(5\mathcal{H}h^0_i+\dot{h}^0_i+\partial_i h^{00}+\delta_{ik}\partial_j h^{jk}-\delta_{jk}\partial_i h^{jk} \right)+\frac{1}{2a^2}\biggl(\partial^k \partial_i h_{0k} + \partial^k\dot{h}_{ik}-\partial^k \partial_k h_{i0}+\\
&-\delta^{jk}\partial_i \dot{h}_{jk}-2\mathcal{H}\delta^k_i \dot{h}_{0k} \biggr);
\label{dG0i}
\end{split}\end{equation}

\begin{equation}\begin{split}
\delta G_{ij}  =& \ddot{a}\left(2ah^{00} \delta_{ij}-\frac{3}{a^3}h_{ij}-ah^{kl}\delta_{kl} \delta_{ij} \right)+a\dot{a}\biggl(\frac{1}{2}\dot{h}^{00}\delta_{ij}-\mathcal{H}h^{00}\delta_{ij}-2\mathcal{H} h^{kl}\delta_{kl}\delta_{ij}-2\mathcal{H}h^{kl}\delta_{ik}\delta_{jl}+\mathcal{H} \delta^{kl} h_{kl} \delta_{ij}+\\
&-\frac{1}{2}\dot{h}^{kl} \delta_{kl} \delta_{ij} \biggr)+\frac{1}{2a^2}\biggl(-\partial_i \dot{h}_{j0}-\partial_j \dot{h}_{i0}+\ddot{h}_{ij}+\partial^k \partial_i h_{jk}+\partial^k\partial_j h_{ki}-\partial^k \partial_k h_{ij}-\partial^k \partial_k h_{00} \delta_{ij}+\partial_i \partial_j h_{00}+\\
&-\delta^{kl}\partial_i \partial_j h_{kl}-\mathcal{H} \dot{h}_{00}\delta_{ij}+\mathcal{H}\delta^{kl}\dot{h}_{kl}\delta_{ij}-2\mathcal{H}\dot{h}_{ij}+2\partial^k \dot{h}_{0k}\delta_{ij}-\delta^{kl}\ddot{h}_{kl} \delta_{ij} -\partial^k \partial^l h_{kl}\delta_{ij} +\delta^{kl} \partial^m \partial_m h_{kl} \delta_{ij}\biggr),
\label{dGij}
\end{split}\end{equation}
where $\mathcal{H}=\dot a/a$ is, as usual, the Hubble parameter.

\subsection{Gauge choice}

It is well-known that, for cosmological perturbations, the split (\ref{metricaperturbata}) into background and perturbation depends on the  choice of coordinates, i.e. on the gauge choice. Metric perturbations are classified in scalar, vector and tensor perturbations, according to their transformation properties under the $SO(3)$ rotation group. In the linear theory, these perturbations are  decoupled and evolve independently one from the other. 
There are $4$ scalar,   $4$ vector and $2$ tensor  independent perturbations. At linear level, tensor perturbations do not couple to matter, whereas vector perturbations decay very fast in an expanding background.   Of the remaining $4$ scalar perturbations,  $2$ are gauge modes, leaving only $2$ physical scalar perturbation modes.

We have two possible choices: either we fix the gauge in order to eliminate the $2$ gauge modes, or we can choose to work with manifest gauge invariant quantities, like e.g. the Bardeen potentials. In this paper, we choose the first approach and use the Newtonian conformal gauge:
\begin{equation}
ds^2 = a^2 \left[-\left(1+2\phi \right)d\eta^2 + \left(1+2\psi \right)\delta_{ij} dx^i dx^j \right].
\end{equation}
In this gauge, using Eqs.~(\ref{dG00}), (\ref{dG0i}) and  Eqs.~(\ref{dGij}), Einstein's field equations give:
\begin{subequations}
\begin{align}
&3\mathcal{H} \left(\mathcal{H}\phi-\dot{\psi} \right)+\nabla^2\psi = 4\pi G a^2 \delta T^0_0;\\
&\partial_i \left(\dot{\psi}-\mathcal{H} \phi \right)=4\pi G a^2 \delta T^0_i;\\
&\left[\left(\mathcal{H}^2+2\dot{\mathcal{H}} \right)\phi+\mathcal{H}\dot{\phi}-\ddot{\psi}-2\mathcal{H}\dot{\psi}\right]\delta^i_j+\frac{1}{2}\left[\nabla^2\left(\psi+\phi \right)\delta^i_j-\partial^i \partial_j \left(\psi+\phi \right) \right] = 4\pi G a^2 \delta T^i_j.
\end{align}
\label{EinsteinNewtonian1}
\end{subequations}
The last equation yields two sets of equations, one for $i \neq j$ and one for the trace.
\subsection{Perturbations of the stress-energy tensor}
\lb{secpst}
We are assuming that our cosmological background  is sourced by a perfect fluid of density $\rho$ and pressure $P$.  We can therefore treat the anisotropic stress  $\pi_{\mu\nu}$ as a perturbation of the stress-energy tensor.  Consistently, we will also treat $\mathcal{S}= 3^{-1/2}\left(p_{\parallel}-p_{\perp}\right)$  as a small perturbation, i.e. we will take $|(p_{\parallel}-p_{\perp})/P|\ll 1$ and consider only terms of order $1$ in the perturbative expansion. 

The background components for the stress-energy tensor are those pertaining to a perfect fluid: $T^0_0 = -\rho,\,T^0_i = T^i_0 = 0,
\,T^i_j = P \delta^i_j$. For the perturbations we have instead 
\be
\delta T^0_0 =-\delta \rho,\quad
\delta T^0_i =-\delta T^i_0 = v_i \left(\rho + P\right),\quad \delta T^i_j = \delta P \delta^i_j + \pi^i_j,
\ee
where $v_i=a\delta u_i$ parametrize fluid velocity perturbations.

Substituting  the previous equations into Eqs.~(\ref{EinsteinNewtonian1}), after some manipulation, we get:
\begin{subequations}
\begin{align}
&3\mathcal{H} \left(\mathcal{H}\phi-\dot{\psi} \right)+\nabla^2\psi = -4\pi G a^2 \delta\rho,\\
&\partial_i \left(\dot{\psi}-\mathcal{H} \phi \right)=4\pi G a^2\left(\rho+P\right)v_i,\\
&\partial^i \partial_j \left(\psi+\phi \right) = -8\pi G a^2 \pi^i_j, \label{ineqjcomponente}\\
&\left(\mathcal{H}^2+2\dot{\mathcal{H}} \right)\phi+\mathcal{H}\dot{\phi}-\ddot{\psi}-2\mathcal{H}\dot{\psi}+\frac{1}{3}\nabla^2\left(\psi+\phi \right) = 4\pi G a^2 \delta P.
\end{align}
\label{EinsteinNewtonian2}
\end{subequations}

The covariant conservation  equation for the stress-energy tensor $\nabla_{\mu} \delta T^{\mu}_{\nu} = 0$ gives two more equations:
\begin{subequations}
\begin{align}
&\dot{\delta \rho}+3\mathcal{H}\left(\delta \rho+\delta P \right)+\left(\rho+P \right)\left(\partial_i v^i+3\dot{\psi} \right)=0;\\
&\left(\rho+P\right)\left(4\mathcal{H}v_i + \dot{v}_i+\partial_i\phi \right)+v_i \left(\dot{\rho}+\dot{P} \right)+\partial_i \delta P+\partial_j \pi^j_i=0.
\end{align}
\lb{kjl}
\end{subequations}

\subsection{dS background: isotropic perturbations}
\lb{secip}
Let us now  consider cosmological perturbations  of the dS background solution, i.e. a background whose   EoS is $P=-\rho$. We first consider the case of isotropic perturbations, i.e. we set the anisotropic stress-tensor $\pi_{ij}=0$ ( equivalently $\mathcal{S}=0)$.  

Using polar coordinates for the 3D spatial sections of the  4D metric  and passing to the corresponding  Fourier space, labeled by wavevector modulus $k$, Eqs.~(\ref{EinsteinNewtonian2})  and   (\ref{kjl}) give  
\begin{subequations}
\begin{align}
&3\mathcal{H} \left(\mathcal{H}\phi_k-\dot{\psi}_k \right)-k^2\psi_k = -4\pi G a^2 \delta \rho_k; \label{isot00}\\
&ik_i \left(\dot{\psi}_k-\mathcal{H}\phi_k \right) =0;\label{isot0i}\\
&-k_ik_j\left(\psi_k+\phi_k\right)=0;\label{isotij}\\
&\left(\mathcal{H}^2+2\dot{\mathcal{H}} \right)\phi_k+\mathcal{H}\dot{\phi}_k-\ddot{\psi}_k-2\mathcal{H}\dot{\psi}_k-\frac{1}{3}k^2 \left(\psi_k+\phi_k\right)=4\pi Ga^2 \delta P_k; \label{isotii}\\
&\dot{\delta \rho}_k+3\mathcal{H} \left(\delta \rho_k+\delta P_k \right)=0;\label{isotcons0}\\
&ik_i \delta P_k=0, \label{isotconsi}
\end{align}
\label{EinsteinNewtonianFourier}
\end{subequations}
 where   quantities with lower index $k$, $\phi_k=\phi_k(k,t)$, $\psi_k=\psi_k(k,t)$, $\delta \rho_k=\delta \rho_k(k,t)$, $\delta P_k=\delta P_k(k,t)$  represent the 3D  Fourier transform of the corresponding quantities. 
Because  Fourier modes  evolve independently, in the following, for sake of simplicity, we will drop the lower index $k$ in the Fourier transforms.

The previous equations can be easily solved.  From Eqs.~ (\ref{isotconsi}) and (\ref{isotij}) we get   $\delta P=0$, $\psi = -\phi$, whereas  Eqs.~(\ref{isot0i}), (\ref{isot00}) and (\ref{isotcons0}) give  $\phi(k, t) = C_{1} (k)a^{-1}$.
Finally,  Eq.~ (\ref{isotcons0}) allows us to determine $\delta\rho$
\be\lb{klu}
\delta \rho = \mathcal{F}(k) a^{-3},  \quad \phi =\mathcal{B}(k) a^{-1}= -\psi =-\frac{4\pi G}{k^2}\mathcal{F}(k) a^{-1},
\ee
where $\mathcal{F}(k)$ and $\mathcal{B}(k)$ are arbitrary functions of $k$, with:
\begin{equation}
\left|\mathcal{F}(k) \right| = \frac{k^2 \left|\mathcal{B}(k) \right|}{4\pi G}.
\end{equation}
This is the well-known relation between the matter density and gravitational potential power  spectra:
\begin{equation}
\langle\left|\mathcal{F}(k) \right|^2 \rangle\propto k^4 \langle\left|\mathcal{B}(k) \right|^2\rangle.
\end{equation}
Our solution  depends on an arbitrary function $\mathcal{F}(k)$ of the Fourier wave vector $k$, therefore it does not determine neither the matter density nor the gravitational potential power spectrum, but only the relation between them. This result does not come unexpected. In fact  Eq.~(\ref {isotconsi}) implies that isotropic perturbations behave as incoherent, stiff, matter: $\delta P=0$.  The dynamics of the perturbations fixes therefore  the EoS, preventing the possibility to impose it from outside. This means that, in the framework we are considering (large-scale regime of a dark energy-dominated universe), the mass distribution at long wavelengths, i.e. the power spectrum (\ref{ps}), cannot be determined by the dynamics  of perturbations. The determination of the power spectrum (\ref{ps}) seems to be only possible in the usual cosmological framework based on inflation, in which the scale-invariant spectrum  is explained in terms of the growing of small perturbations generated in the early universe. Physically, this expresses the fact   that the large-scale distribution of matter cannot be determined by the interaction between dark energy and baryonic matter, the latter being relevant for the distribution at small scales only.
Thus, in our approach  the observed long-wavelength power spectrum (\ref{ps}) has to be  used to determine the arbitrary function $\mathcal{F}(k)$. Assuming the validity of Eq.~(\ref{ps}), we get $\langle\left|\mathcal{B}(k) \right|^2\rangle \sim k^{-3}$. 

\subsection{dS background: anisotropic perturbations}
\label{dsan}
Let us now pass to  consider anisotropic perturbations of the dS background, i.e. the case $\pi_{ij}\neq0$. Taking into account the considerations of  sections \ref{secset} and \ref{secpst}, this boils down to consider perturbations generated by an anisotropic fluid with  $p_{\parallel} \neq p_{\perp}$.  We are now dealing with the small-scale regime of our cosmological model,   for which we expect the spatial distribution of inhomogeneities to be determined by the dynamics of perturbations. Thus,  the distribution of small-scale structures in our universe is determined  by our effective anisotropic fluid, which encodes the interaction between   dark energy and baryonic matter.

Eqs.~(\ref{EinsteinNewtonian2})  and   (\ref{kjl}) now give:
\begin{subequations}
\begin{align}
&3\mathcal{H} \left(\mathcal{H}\phi-\dot{\psi} \right)+\nabla^2\psi = -4\pi G a^2 \delta\rho; \label{00}\\
&\partial_i \left(\dot{\psi}-\mathcal{H} \phi \right)=0;\label{0i} \\
&\partial_i \partial_j \left(\psi+\phi\right) = -8\pi G a^2 \pi_{ij}; \label{ij}\\
&\left(\mathcal{H}^2+2\dot{\mathcal{H}} \right)\phi+\mathcal{H}\dot{\phi}-\ddot{\psi}-2\mathcal{H}\dot{\psi}+\frac{1}{3}\nabla^2\left(\psi+\phi \right) = 4\pi G a^2 \delta P; \label{ii}\\
&\dot{\delta \rho}+3\mathcal{H}\left(\delta \rho+\delta P \right)=0 ; \label{conservation0}\\
&\partial_i \delta P+\partial_j \pi^j_i=0. \label{conservationi}
\end{align}
\label{EinsteinNewtonian3}
\end{subequations}
We consider anisotropic perturbations that can be derived by a scalar potential 
$\Pi$. Being $\pi_{ij}$ traceless, we can write:
\begin{equation}
\pi_{ij} \equiv \partial_i \partial_j \Pi-\frac{1}{3}\delta_{ij}\nabla^2 \Pi.
\end{equation}

This allows us to simplify drastically our system of equations. Passing to Fourier space and dropping  the lower index in the Fourier transforms in order to simplify the notation, from Eqs.~(\ref{00}), (\ref{0i}), (\ref{ij}) and (\ref{conservationi}) we get:
\be\lb{klt}
\phi + \psi = -8\pi Ga^2 \Pi,\quad k^2\psi = 4\pi G a^2 \delta \rho, \quad \delta P=\frac{2}{3}k^2 \Pi.
\ee
The last equation above is fully consistent with the fact that the anisotropy in the perturbation is linked directly to $\delta P\neq0$, as we have seen in Sect. \ref{secpst}.

The only other independent equation in  the system is the conservation equation (\ref{conservation0}), which can be written  as:
\be\lb{bhn}
\dot{\delta \rho}+3\mathcal{H} (\delta \rho+\delta P) =0.
\ee 
The other equations  become identities after using Eqs.~(\ref{conservationi}) and  (\ref{bhn}).
We are therefore left with a system of $4$ equations in  $5$ unknowns. As expected, we need an equation of state for the perturbations in order to close the system.
As discussed in Sect. \ref{seceos},  this information is encoded in Eq.~(\ref{eeos}), which is inherited  from the galactic dynamics. 

Since we are considering small perturbations of the dS spacetime due to an anisotropic  fluid, the pressure perturbation $\delta P$  in Eq.~(\ref{bhn}) can be identified with the dark force  (\ref{eeos}), i.e. $\delta P=p_{\parallel}$.  This identification is evident from  Eq.~(\ref{pparallel}), which allows to  write $p_{\parallel}$ as a background pressure plus the anisotropic stress contribution.  Furthermore, in Eq.~(\ref{eeos}), the effective matter density $\rho_E$ is the source of the dark force. We can therefore set $\delta \rho=\rho_E$ in Eq.~(\ref{bhn}).
Eq.~(\ref{eeos}) determines only the spatial profile of $\delta P$ once $\delta \rho$ is known, whereas it is insensitive to their dependence on the conformal time.  This is consistent with the  fact that Eq.~(\ref{eeos}) is originated  in galactic dynamics. 

In order to solve the system, we therefore need a factorization  of $\delta P$ and $\delta \rho$ in space- and time-profiles and we also need  an EoS consistent  with this factorization:
\be\lb{jhk}
 \delta P=w \delta \rho ,\quad \delta \rho(r, t) = \delta \hat{\rho}(t) \delta \rho(r), \quad \delta P(r, t) = \delta \hat{P}(t) \delta P(r),
\ee
 with $w$ constant. Notice that we are using a perfect fluid equation of state for the perturbation.   
Since Eq.~(\ref{eeos}) is written in  terms of the radial coordinate $r$, we will also solve Eq.~(\ref{bhn}) in coordinate space.    

Differentiating (\ref{eeos}) with respect to $r$ and using Eq.~(\ref{jhk}), we get:
\be\lb{oop}
 \delta \dot{\hat{\rho}}+3\mathcal{H}(1+w) \delta \hat{\rho} = 0,\quad \frac{d }{dr}\left[r^2 \delta \rho\right]=\frac{H_0}{16\pi G w^2}.  
\ee
It is quite interesting  to notice that our ansatz  (\ref{jhk})  allows to perform, at least at perturbative level, the same decoupling of cosmological degrees of freedom from inhomegenities  we have described in Sect. \ref{sec3} in our anisotropic fluid cosmology. The EoS, $ \delta P=w \delta \rho$, determines, through the first equation in (\ref{oop}), the time-dependence of the homogenous part of the matter density, whereas  Eq.~(\ref{eeos}) determines the inhomogeneity profile trough the second equation in (\ref{oop}).

The general solution of the second equation in    (\ref{oop}) contains a term proportional to $1/r$ and a term  $\beta/r^2$, with $\beta$ integration constant. Eq.~(\ref{eeos}) requires $\beta=0$ so that the solution of (\ref{oop}) is: 
\be\lb{ggg}
\delta \hat{\rho}(t) \sim a^{-3(1+w)}, \quad \delta {\rho}(r)= \frac{H_0}{16\pi G w^2}\frac{1}{r}.
\ee
The cosmological evolution of the homogeneous part of the perturbation is that pertaining  to a perfect fluid, whereas the profile for inhomogeneities is given by  an harmonic function in 3D.

The Fourier transform of the spatial profile of $\rho$   gives $\delta \rho_k\sim \frac{H_0}{16\pi G w^2}\frac{1}{k^2}$ and the power spectrum is:
\begin{equation}\lb{hhj}
P(k) = \langle \delta \rho_k^2 \rangle = \frac{\int d^3 k \ \delta \rho_k^2}{\int d^3 k}\sim \left(\frac{H_0}{16\pi G w^2}\right)^2\frac{1}{k^4}.
\end{equation}
This is  the result of our cosmological model for the power spectrum of mass distribution at short wavelengths. It gives a theoretical determination of the transfer function $T(k)$ in Eq.~ (\ref{ps1}).\\
Since our model deals with the late-time cosmology, describing the phenomenology of dark energy, baryonic matter and their interaction, it does not come as a surprise we are only able to reproduce  the power spectrum (\ref{ps2}), predicted by the galaxy two-point correlation function. On the other hand it  fails to reproduce the power spectrum at the equivalence epoch Eq.~(\ref{ps1.2}), which depends on the physics governing  matter radiation at the equivalence epoch.

\section{Conclusions}
\label{concl}
In this paper we have proposed an anisotropic fluid  cosmological model for describing  our present,  dark energy-dominated, universe. 
The model does not assume the presence of dark matter.  Dark energy, baryonic matter and their possible  effective interaction are  codified in a peculiar EoS for the anisotropic fluid. This EoS  is inherited from that used to explain galactic dynamics without assuming the presence of dark matter \cite{Cadoni:2017evg}.

We have shown that the  model has several nice features.  The anisotropy in the fluid pressure allows to generate inhomogeneities at small scales in a natural way. It can be therefore used  to explain mass distribution, i.e the matter density power spectrum, at short wavelengths. Cosmological dynamics, i.e. time evolution for the scale factor and for the homogeneous component of matter density, completely decouples from inhomogeneities. The former is  ruled  by usual FLRW cosmology, whereas the latter are determined by the relation between pressure and density of the anisotropic fluid. 

We have also found that the predictions of our model concerning  the accelerated expansion of the universe and mass distribution at small scales are in accordance with observations. In the large  distances regime, our model is well-described by a generalized Chaplygin gas  and fits  observational data from type IA supernovae,  used to probe the present  accelerated expansion of our universe.
In the short distance regime, we have used perturbation theory near the dS background to describe mass distribution. Perturbations due to the anisotropic fluid are described by an anisotropic stress tensor. We find a power spectrum  $P(k)$ for mass density distribution at short wavelengths behaving as $1/k^4$, in good accordance with the observed  2-point galaxy correlation function {for matter distribution at small scales.

Let us conclude with the drawbacks of our  approach. In the present form, our anisotropic fluid cosmological model  can be only used to describe a dark energy-dominated universe, but not to describe the early-time cosmology and the radiation/baryonic matter dominated eras. This is because any mass distribution is intrinsically unstable in FLRW cosmology. Therefore, any inhomogeneous, FLRW-based cosmological model can only be used to describe late-time cosmology and  cosmic structures at small scales, but not the evolution of perturbations from early-time cosmology. Thus, large scale structures can only be explained in the framework of usual FLRW cosmology, in terms of the growing of small perturbations of the early universe described by linear perturbation theory. 

This is fully consistent with the  results of our paper. We have seen in Sect. \ref{secip} that the behaviour of the power spectrum $P(k)$ for the mass density distribution, at large wavelengths, is not determined by perturbations in the present dark energy-dominated universe. The observed linear scaling  $P(k)\sim k$  has to be explained in terms of small fluctuations in the early universe.
Conversely,  small scale structures find a natural explanation in our model as inhomogeneities triggered by an anisotropic stress tensor. This is in turn consistent with the fact that the presence of dark matter is crucial for structure formation in the $\Lambda$CDM model. As expected, in our cosmological model the anisotropic stress tensor, generated by the assumed dark energy/baryonic matter interaction, plays the same role that dark matter plays in the $\Lambda$CDM model for small scale structure formations.

It is an open question whether some alternative description of anisotropic fluid cosmology, not based on the inhomogeneous metric (\ref{seccosm}), could exist.
In this paper we have generated a  non-trivial anisotropic fluid by allowing for a dependence of $p_{\perp}, \ p_{\parallel}$ and $\rho$ from the radial coordinate $r$ (see the discussion around Eq.~(\ref{seccosm})).  
This has the advantage of  linking directly the anisotropies  in the stress tensor to inhomogeneities, but prevents the use of the model for early cosmology.
It is possible that relaxing this condition could make anisotropic fluid cosmology suitable also for describing early-time cosmology. The results  of Refs. \cite{Brevik:2017msy, Kolekar:2019oqj} where anisotropic fluid cosmology have been used to describe inflation, seem to support this point of view.
\section*{Acknowledgements}
This research was partially supported by INFN, research initiative 
QUAGRAP (M.C.).

\end{document}